\documentclass[prb,twocolumn,showpacs,preprintnumbers,amsmath,amssymb]{revtex4}

\usepackage{graphicx}
\usepackage{dcolumn}
\usepackage{bm}

\usepackage{color}

\begin{document}


\title{
The Kondo crossover in shot noise of a single quantum dot \\
with orbital degeneracy
}

\author{R. Sakano$^1$}
\author{T. Fujii$^2$}%
\author{A. Oguri$^3$}

\affiliation{%
$^1$Department of Applied Physics, University of Tokyo, Bunkyo, Tokyo, Japan\\
$^2$Institute for Solid State Physics, University of Tokyo, Kashiwa, Chiba, Japan\\
$^3$Department of Physics, Osaka City University, Sumiyoshi, Osaka, Japan
}%

\date{\today}

\begin{abstract}
We investigate out of equilibrium transport through an orbital Kondo system 
realized in a single quantum dot, 
described by the multiorbital impurity Anderson model. 
Shot noise and current are calculated up to 
the third order in bias voltage 
in the particle-hole symmetric case,
using the renormalized perturbation theory.
The derived expressions are asymptotically exact 
at low energies.
The resulting Fano factor of the backscattering current $F_b$
is expressed in terms of
the Wilson ratio $R$ and the orbital degeneracy $N$ as 
$F_b =\frac{1 + 9(N-1)(R-1)^2}{1 + 5(N-1)(R-1)^2}$ at zero temperature.
Then, for small Coulomb repulsions $U$,
we calculate the Fano factor
exactly up to terms of order $U^5$,
and also carry out the numerical renormalization group calculation 
for intermediate $U$ in the case of two fold and four fold
degeneracy ($N=2,\,4$).
As $U$ increases, the charge fluctuation in the dot is suppressed,
and the Fano factor varies rapidly from the noninteracting value $F_b=1$
to the value in the 
Kondo limit $F_b=\frac{N+8}{N+4}$,
near the crossover region $U\sim \pi \Gamma$,
with the energy scale of the hybridization $\Gamma$.
\end{abstract}

\maketitle

\section{introduction}

Kondo physics has been studied since the 1960s 
for dilute magnetic alloys and heavy fermion systems
\cite{HewsonBook}.
Advancements in nanofabrication techniques in these years 
have made it possible to experimentally achieve the nonequilibrium 
Kondo state in quantum dots by applying a bias voltage, 
opening a new paradigm of Kondo physics. 
In the early days most research on the nonequilibrium Kondo effect 
has focused
on the time averaged current or linear conductance  
\cite{PhysRevLett.100.246601,PhysRevB.79.165413}.
Recently, the shot noise associated with the Kondo effect 
has attracted the
attention of the researchers in this field
\cite{
PhysRevB.58.14978,PhysRevLett.88.116802,0953-8984-14-19-318,PhysRevLett.90.116602,PhysRevB.73.233310,PhysRevB.73.195301,PhysRevB.81.241305}.

Shot noise measurements in mesoscopic devices provide 
important information about 
the effective charge $e^*$ of current-carrying particles.
For instance, the fractional charge $e^*=e/3$ of 
a fractional quantum Hall system has been clarified through 
the shot noise  measurement, 
and also the charge  $e^*=2e$ of the Cooper-pair 
has been observed in 
normal metal/superconductor junctions 
\cite{Nature.389.162,PhysRevLett.79.2526,PhysRevLett.90.067002}.
Furthermore, for other correlated 
electron systems,
the shot noise has
become
an important probe to study 
the properties of the low-energy excitations.
In quantum dot systems,
the observation of a fractional enhancement of $5e/3$
for the {\it backscattering current\/} in a symmetric barrier 
\cite{PhysRevLett.97.016602,PhysRevLett.97.086601} has stimulated 
recent studies of the shot noise
in the Kondo regime
\cite{PhysRevB.77.241303,PCkobayashi}.
Note that the {\it backscattering current\/} is
an essential quantity for observing the shot noise,
and is defined by the deviation
from the value of the linear current in the unitarity limit.

Theories for the shot noise in
Kondo systems have been extended,
naturally, to another class of exotic Kondo system
: the so-called orbital Kondo effect which  has been observed in experiments
\cite{PhysRevLett.93.017205,Nature434.484,PhysRevB.75.241407,PhysRevLett.87.216803,
PhysRevLett.90.026602,PhysRevLett.94.186406,JPSJ.74.95,PhysRevLett.95.067204,
PhysRevB.73.155332,PhysRevB.73.241305,PhysRevB.74.205119,JPSJ.77.094707,PhysRevB.80.155330}.
In the case of multiple quantum dots or a single dot with a symmetrical shape, 
the system can have
orbital degeneracy.
The orbital degrees of freedom have been expected to affect
significantly the low-energy properties,
which can still be described by the local Fermi-liquid theory
\cite{PTP.55.67}. 
The shot noise for systems with orbital degeneracy 
has been investigated in the Kondo limit,
where the local charge degrees of freedom in the dot site are quenched
\cite{PhysRevLett.100.036603,PhysRevLett.100.036604,PhysRevB.80.125304,PhysRevB.80.155322}.
These studies have provided the value of the Fano factor of the backscattering current which depends on the orbital degeneracy $N$,
and have
inspired experimental noise measurements in a single-wall carbon nanotube
\cite{NaturePhys5.208}. 

There have been some qualitative discussions based on
the Anderson model that capture the physics away 
from the Kondo limit through the local charge degrees of freedom
that remain
active for finite on-site 
Coulomb repulsions $U$
\cite{PhysRevB.81.115327}. 
The approximations used, however, were not applicable to 
low energies, and
reliable results in the low-temperature Fermi-liquid regime
are desired.
In a previous work, Fujii has derived the expression 
of the Fano factor for the single orbital Anderson model ($N=2$)  
in the particle-hole symmetric case 
using the renormalized perturbation theory (RPT) \cite{JPhysSocJpn.79.044714}. 
The RPT is an approach that starts with the Fermi-liquid ground state, 
and gives exact asymptotic behavior of the correlation functions 
 at low frequencies, low temperatures, and low bias voltages 
\cite{PhysRevLett.70.4007,0953-8984-5-34-014,PhysRevB.64.153305,JPhysCondMatt.17.5413}. 
Therefore, 
Fujii's result for the Fano factor $F_b$ 
is asymptotically exact at low energies,   
and is expressed in terms of a single parameter as  
$F_b =\frac{1 + 9(R-1)^2}{1 + 5(R-1)^2}$, 
where $R$ is  the Wilson ratio that determines 
the universal Kondo behavior 
for all values of the Coulomb 
repulsion $0\leq U<\infty$ \cite{JPhysSocJpn.79.044714}.

The exact result for the Fano factor for the orbital Kondo system
with $N$ orbitals has already been given for the Kondo limit ($U \to \infty$), 
as mentioned above. 
Away from the Kondo regime,  however, 
it still has  not been clarified as to how the Fano factor varies 
with the value of the Coulomb repulsion. 
The purpose of the present paper is to provide 
the exact low-energy expression of the Fano factor for arbitrary $N$ and $R$,
on the basis of the multiorbital impurity Anderson model.
To this end,  following the derivation in the single orbital case, 
we use the RPT and consider the particle-hole symmetric case
 \cite{JPhysSocJpn.79.044714}.
Specifically, we start with a general formulation for the shot noise,  
which is based on the nonequilibrium Kubo formula 
for mesoscopic systems 
\cite{JPhysSocJpn.76.044709},
and give the expression of the Fano factor 
in terms of the Wilson ratio $R$ and the orbital degeneracy $N$.
Equations \eqref{eq:shotnoise_fromC2} and \eqref{eq:fano} are the main results of the paper.
We also calculate the explicit form of the Fano factor
for small Coulomb repulsions up to terms of order $U^5$,
and for intermediate $U$ we 
carry out the numerical renormalization group (NRG) calculations 
for the system with $N=2$ and $N=4$.
The Fano factor 
varies rapidly near $U\sim \pi \Gamma$ 
from the noninteracting value $F_b=1$  
to the value $F_b = (N+8)/(N+4)$ in the Kondo limit,
as the crossover from the weak coupling regime 
to the Kondo regime takes place.
Here, $\Gamma$ is the linewidth of the dot levels 
owing to the coupling to the leads.

This paper is organized as follows.
In Sec.\ \ref{sec:model&cal}, we introduce a generalized impurity Anderson model for the multiorbital quantum dot system, 
and give a brief explanation for the RPT
to describe the low-energy states.
In Sec.\  \ref{sec:transport}, 
we describe the derivation of the exact expression for 
the current and the shot noise at low energies.
Then, we discuss the $U$ dependence of the Fano factor 
calculated with the NRG.
A brief discussion and summary are given in Sec.\ \ref{sec:summary}.

\section{Model and calculation}\label{sec:model&cal}

\subsection{Multiorbital impurity Anderson model}

\begin{figure}[bt]
\includegraphics[width=4cm]{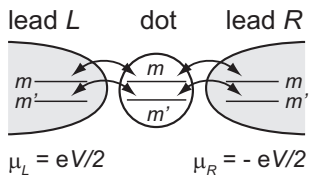}
\caption{\label{fig:schematic}
A schematic plot of the single quantum dot system with orbital degrees of freedom.
The dot is connected to two leads via electron tunneling, 
conserving the orbital quantum number.}
\end{figure}

Let us consider a single quantum dot system with $N$-degenerate orbitals. 
We assume
the orbital conservation in the tunneling process between the dot and the leads in
the manner shown in Fig.\ \ref{fig:schematic}.
This assumption has been experimentally confirmed to be reasonable
for vertical quantum dot and carbon nanotube quantum dot systems \cite{PhysRevLett.93.017205,Nature434.484}.
In these assumptions, our system can be described by 
a multiorbital version of the impurity Anderson model,
\begin{eqnarray}
{\cal H} &=& {\cal H}_0 + {\cal H}_1, \label{eq:mia-model} \\
{\cal H}_0 &=& \sum_{k\alpha m} \varepsilon_{k \alpha m} c^{\dagger}_{k \alpha m} c_{k \alpha m}^{}
+ \sum_m \epsilon_{d m} d_{m}^{\dagger} d_{m}^{} \\
\nonumber
&& \qquad + \sum_{k\alpha m} \left( V_{k \alpha m} c^{\dagger}_{k\alpha m} d_{m}^{} + \mbox{h.c.} \right), \\
{\cal H}_1 &=& \frac{1}{2} \sum_{m \neq m'} U d_{m}^{\dagger} d_{m}^{} d_{m'}^{\dagger} d_{m'}^{}, \label{eq:AndersonHamiltonian}
\end{eqnarray}
where $d_m (d_m^{\dagger})$ annihilates (creates) an electron in the dot level $\epsilon_{dm}$ with state $m$,
$c_{k \alpha m}^{} (c_{k\alpha m}^{\dagger})$ annihilates (creates) a conduction electron with momentum $k$ and state $m$ in the lead $\alpha = L,R$,
and $U$ is the on-site Coulomb repulsion in the quantum dot.
Here, $m$ indicates for
the $N$-fold orbital state ($m=1,2, \cdots, N$).
To investigate the particle-hole symmetric Kondo effect,
the dot level is chosen to be  $\epsilon_{dm} = - (N-1)U/2$ 
and the degeneracy $N$ to be an even number.
The intrinsic linewidth of the dot levels owing to tunnel coupling $V_{k\alpha m}$ is
$\Gamma(\omega) = (\Gamma_L(\omega) + \Gamma_R (\omega) )/2$
with
$\Gamma_{\alpha}(\omega) = 2\pi \sum_{k} \delta(\omega - \varepsilon_{k \alpha m}) |V_{k \alpha m}|^2$,
which is assumed to be
independent of the orbital state $m$.
For conduction electrons, without any prominent features,
$\Gamma_{\alpha} (\omega)$ does not have a strong dependence on $\omega$,
so it is usual to take the case of
a wide conduction band
with a flat density of states limit,
where $\Gamma_{\alpha} (\omega)$ can be taken as a constant $\Gamma_{\alpha}$.
It has been shown that
an asymmetric barrier gives rise to 
a non-universal shift in 
the fractional enhancement for the backscattering current
\cite{PhysRevB.80.155322}. 
To discuss the shot noise properties originating from the Kondo correlation, 
 the lead-dot couplings are assumed to be symmetric:
$\Gamma = \Gamma_L = \Gamma_R$.
The chemical potentials $\mu_{L/R}= \pm eV/2$,
satisfying $\mu_L - \mu_R = eV$,
are measured relative to the Fermi level which is defined 
at zero voltage $V=0$ such that $\mu_L=\mu_R=0$.

\subsection{Fermi-liquid description for low-energy states}

We make use of the renormalized perturbation theory
\cite{FieldTheoryBook},
which has been successfully applied to the impurity Anderson model
in equilibrium
\cite{PhysRevLett.70.4007,0953-8984-5-34-014},
and later extended to low-bias steady states
\cite{JPSJ.74.110, JPhysCondMatt.17.5413, JPhysSocJpn.79.044714}.
In this section,
we outline the approach to 
the multiorbital impurity Anderson model 
described by Eq.\ (\ref{eq:mia-model}). 
Furthermore, 
we provide explicit expressions 
for the renormalized parameters for small $U$ 
obtained exactly up to three-loop contributions for general $N$.

The three basic parameters that specify the model are the energy level of the dot state $\epsilon_{dm}$, the linewidth parameter $\Gamma$ and the Coulomb repulsion $U$ in the dot.
The low-energy properties of the system can be characterized by 
the quasiparticles with the renormalized parameters, 
defined by 
\begin{eqnarray}
\bar{\epsilon}_{dm} &=& z \left( \epsilon_{dm} + \Sigma_{dm}^r (0) \right) \;,
\label{eq:ed_ren}
\\
\bar{\Gamma} &=& z \Gamma \;,
\label{eq:Gam_ren}
\\
\bar{U} &=& z^2 \Gamma_{mm'}^{(4)}(0,0,0,0) \quad (m \neq m') \;,
\label{eq:U_ren}
\end{eqnarray}
where $\Sigma_{dm}^r (\omega)$ is the self-energy of the retarded Green function
for the dot state: 
$G_{dm}^r (\omega) = [ \omega - \epsilon_{dm}^{} + i \Gamma - \Sigma_{dm}^r (\omega) ]^{-1}$,
$z = [ 1 - \partial \Sigma_{dm}^{r}(\omega)/\partial \omega|_{\omega=0} ]^{-1}$ is the wave function renormalization factor,
and $\Gamma_{mm'}^{(4)}(\omega_1, \omega_2, \omega_3, \omega_4)$ is the local full four-point vertex function for the scattering of the electrons 
with the orbital $m$ and $m'$.
The perturbation theory in powers of $U$ can be reorganized as 
an expansion with respect to the renormalized interaction $\bar{U}$, 
taking the free quasiparticle
Green's function $\bar{g}_{dm}^r (\omega) = ( \omega - \bar{\epsilon}_{dm} + i \bar{\Gamma})^{-1}$ as the zero-order propagator
\cite{PTP46.244,PhysRevLett.67.3720,PhysRevB.68.155310}.
The three counter terms have to be included to prevent 
any further renormalization of the parameters, 
$\bar{\epsilon}_{dm}$, $\bar{\Gamma}$, and $\bar{U}$.
Then, the low energy properties can be specified entirely 
in terms of the three renormalized parameters. 
These procedures are the
basis
of the renormalized perturbation for the impurity Anderson model,
presented by Hewson
\cite{PhysRevLett.70.4007}.

At low bias voltages described by the Fermi-liquid theory, 
there are explicit
relations between the renormalized parameters and the susceptibilities,
\begin{eqnarray}
\frac{\bar{U}}{\pi \bar{\Gamma}} &=& \frac{ R-1}{\sin^2(\pi n_{dm}^{})} \;, 
\label{eq:Uren2}
\\
\bar{\Gamma} &=& \frac{N}{(N-1)\chi_{d}^{\ast} + \chi_{c,d}^{\ast}} \, \Gamma  
\;,
\label{eq:Gam_ren2}
\\
\bar{\epsilon}_d^{} &=&  \bar{\Gamma}\,  \cot \left(\pi n_{dm}^{}\right)
\label{eq:ed_ren2}
\;.
\end{eqnarray}
Here, $n_{dm}$ is the average number of electrons in the dot-site,
$R$ is the Wilson ratio defined by
\begin{eqnarray}
R \,\equiv \,
\frac{\chi_{d}^{\ast}}{\gamma_d^{\ast}} \, 
= \frac{N}{(N-1) + \chi_{c,d}^{\ast}/\chi_{d}^{\ast}}\;, 
\label{eq:WR}
\end{eqnarray}
where
$\chi_{d}^{\ast}= 1- \partial \Sigma_{dm}^{r}(\omega)/\partial h |_{\omega=0,h=0}^{}$
with the Zeeman energy at the dot-site $h$ is the enhancement factor for the dot susceptibility,
$\chi_{c,d}^{\ast}= 1 + \partial \Sigma_{dm}^{r}(\omega)/\partial \epsilon_d |_{\omega=0}^{}$
is the enhancement factor for the charge
susceptibility,
and $\gamma_d^{\ast} \equiv z^{-1}$ is the enhancement factor 
for the $T$-linear specific heat coefficient.

In the particle-hole symmetric case,
 the electron filling is given by $n_{dm}=1/2$, 
and  the relations above can be simplified as,
\begin{eqnarray}
\frac{\bar{U}}{\pi \bar{\Gamma}}=R-1, 
\qquad \quad 
 \bar{\epsilon}_{dm} = 0\;. \label{eq:relation-symmetric}
\end{eqnarray}
In the Kondo limit ($U \to \infty$),
the charge susceptibility is suppressed $\chi_{c,d}^{\ast} \to 0$,
and $\bar{\Gamma}$ can be considered as the Kondo temperature as,
$T_K = \pi \bar{\Gamma}/4$.
Therefore, Eq.\ (\ref{eq:WR}) shows that 
the Wilson ratio converges to $R \to N/(N-1)$ in the Kondo limit.
In the noninteracting case ($U=0$), 
the parameters take the value 
$\chi_{d}^{\ast} = \chi_{c,d}^{\ast} = \gamma_{d}^{\ast}=1$ by definition.

The behavior of the renormalized parameters 
for small $U$ can be clarified with the perturbation approach. 
We have calculated the enhancement factor $\gamma_d^*$ 
and the four-point vertex for $m \neq m'$ 
in the particle-hole symmetric case,
extending the calculations of Yamada-Yosida
for $N=2$ \cite{PTP46.244,0953-8984-13-44-314} to general $N$,
\begin{widetext}
\begin{eqnarray}
&& \gamma_d^* 
= 1+ \left( 3-\frac{\pi^2}{4}\right)(N-1)\, u^2 
-\left( \frac{21}{2} \zeta(3) - 7 - \frac{\pi^2}{2} \right)(N-1)(N-2)\, u^3 
+ {\cal O}(u^4)
\;, \label{eq:enhancementfactor-ex} \\
&&
\frac{1}{\pi \Gamma}
\Gamma_{mm'}^{(4)}(0,0,0,0)
= u
- (N-2) u^2
+ \biggl[ N^2+\left(1-\frac{\pi ^2}{2}\right) N-\frac{\pi^2}{2}+9 \biggr] u^3 \nonumber \\
&& \qquad\qquad\qquad\qquad\quad - (N-2) \left[ N^2 + \left(21 \zeta(3) - \frac{7}{4} \pi^2 - 12 \right) N + \frac{133}{2} \zeta (3) - \frac{71}{12} \pi^2 - 17 \right] u^4
+ {\cal O}(u^5) \;, \label{eq:forvertex-ex}
\end{eqnarray}
where $u \equiv U/(\pi\Gamma)$ 
and $\zeta(x)$ is the Riemann zeta function.
Note that $\gamma_d^*$ captures
the terms of odd order
in $U$ for $N>2$.
Correspondingly, $\Gamma_{mm'}^{(4)}(0,0,0,0)$ has 
finite contributions from
the terms of even order
for $N>2$.
The appearance of the zeta function, 
which is absent in the case of $N=2$,
in the coefficients of the perturbation series
is also caused by the orbital degeneracy.
Substituting Eqs.\ \eqref{eq:enhancementfactor-ex} and \eqref{eq:forvertex-ex}
into Eqs.\ \eqref{eq:Gam_ren} and \eqref{eq:U_ren}, 
and then through Eq.\ \eqref{eq:relation-symmetric}, 
we have 
obtained the Wilson ratio exactly up to terms of order $U^4$, 
\begin{eqnarray}
R &=&
1
+u
-(N-2) u^2
+ \biggl[ N^2-\left( 2 + \frac{\pi^2}{4} \right) N - \frac{3}{4} \pi^2 + 12 \biggr] u^3 \nonumber \\
&& \quad - (N-2) \biggl[ N^2 + \left( \frac{21}{2} \zeta (3) - \pi^2 - 8 \right)N + 77 \zeta (3) - \frac{20}{3} \pi ^2 - 21 \biggr] u^4
+ {\cal O}(u^5)\;. \label{eq:WR-ex}
\end{eqnarray}
\end{widetext}
We note
that $R$ is an alternating series  up to order $U^4$ for $N \geq 4$.

For intermediate values of $U$, 
the renormalized parameters can be calculated with the NRG approach 
in the case of $N=2$ and $4$, and
the Bethe ansatz exact solution (BAE)
in the case of $N=2$
\cite{RevModPhys.47.773,PhysRevB.21.1003,Wiegmann1980163,Kawakami1981483}.

\section{
Renormalized perturbation theory for nonequilibrium transport
}\label{sec:transport}


Here, we derive the expression for the current and shot noise
of the multiorbital impurity Anderson model
using the second-order perturbation in the renormalized interaction $\bar{U}$.
The obtained expression covers not only the strong-coupling
\cite{PhysRevLett.97.086601,PhysRevLett.97.016602,PhysRevB.73.233310,PhysRevLett.100.036604,PhysRevLett.100.036603}
and weak-coupling limit
\cite{PhysRevLett.67.3720,PhysRevB.73.195301}
but also the whole range of the Coulomb repulsion $U$,
which is an extention of the previous SU(2) result
\cite{PhysRevB.64.153305,JPhysSocJpn.79.044714}
to general $N$.
One of the significant advantages of the RPT is that
the current and shot noise calculated
up to the second order in $\bar{U}$ are asymptotically exact at low energies,
up to the third order in the bias voltage.
From the results for the shot noise,
we also derive the exact Fano factor of the backscattering current
for low bias voltages.

\subsection{Current}
The symmetrized current operator $J$ across the quantum dot system described by Eq.\ (\ref{eq:AndersonHamiltonian}) is written as,
\begin{eqnarray}
J = \frac{J_L - J_R}{2} \label{eq:current-operator},
\end{eqnarray}
where,
\begin{eqnarray}
J_{\alpha} &=& -e \frac{d}{dt} N_{\alpha} \nonumber \\
&=& i\frac{e}{h} \sum_{km} \left( V_{k \alpha m} c_{k \alpha m}^{\dagger} d_{m}^{} - V_{k\alpha m}^* d_{m}^{\dagger} c_{k \alpha m}^{} \right) \;, \nonumber 
\\
\end{eqnarray}
is the current operator for the electrons tunneling from lead $\alpha$
to the quantum dot with number operator for electrons in lead $\alpha$,
$N_{\alpha} = \sum_{km} c_{k\alpha m}^{\dagger} c_{k\alpha m}^{}$.
A general formula for the time-averaged current $I$ owing to the applied voltage through a quantum dot has been given by Hershfield {\it et al.}
\cite{PhysRevLett.67.3720,PhysRevB.46.7046}
and Meir and Wingreen
\cite{PhysRevLett.68.2512, PhysRevB.49.11040}. 
The formula can be specialized for the symmetric lead-dot coupling as,
\begin{eqnarray}
I = \left\langle J \right\rangle = \frac{e}{h} \sum_m \int d\omega \, T_m (\omega) \left( f_L(\omega) - f_R(\omega) \right)\;,
\label{eq:current_formula}
\end{eqnarray}
where 
$T_m(\omega) = - \Gamma\, \mbox{Im} G^r_d (\omega)$ is 
the transmission probability for a channel $m$,
and 
$f_{\alpha}(\omega) = \left[ \exp((\omega - \mu_{\alpha})/T) +1 \right]^{-1}$
is the Fermi distribution function for the electrons in lead $\alpha$.
Here, the average
$\langle \cdots \rangle$
is taken over the density matrix
\cite{PhysRevLett.70.2134,JPSJ.74.110}.
In the practical calculation,
we make use of the density matrix for the nonequilibrium steady state.
The renormalized dot self-energy can be calculated exactly 
up to order $\omega^2, V^2$, and $T^2$,
extending the calculations of Ref. \onlinecite{PhysRevB.64.153305}
for $N=2$ to general $N$.
In the particle-hole symmetric case, it takes the form,
\begin{eqnarray}
\bar{\Sigma}_{dm}^r (\omega) = -i\frac{(N-1)}{2\bar{\Gamma}} \left(\frac{\bar{U}}{\pi \bar{\Gamma}} \right)^2
\left[ \omega^2 + \frac{3}{4}(eV)^2 + (\pi T)^2 \right]. \nonumber \\
\label{eq:rpt_self_2}
\end{eqnarray}
From this result, $T_m(\omega)$ can also be determined 
exactly up to order $\omega^2, (eV)^2$ and $T^2$,
which leads to the time-averaged current at $T=0$ up to $V^3$ 
from Eq.\ \eqref{eq:current_formula}, 
\begin{eqnarray}
I
\,=\, \frac{Ne^2}{h}V \left[ 1 
- 
\frac{1+5(N-1) (R-1)^2}{12} \left( \frac{eV}{\bar{\Gamma}} \right)^2 \right] . \nonumber \\
\label{eq:current}
\end{eqnarray}
For $N=2$, this expression corresponds to Oguri's result 
for the symmetric spin Anderson model
\cite{PhysRevB.64.153305,JPSJ.71.2969,JPSJ.74.110}.
The time-averaged current in the Kondo limit
also agrees with Mora {\it et al.}'s result
for the half-filled and symmetric lead-dot coupling case
\cite{PhysRevB.80.155322}.

\subsection{Shot noise}

The current noise $S$ is defined by 
the correlation function for the current fluctuation,
\begin{eqnarray}
S \equiv \int dt \left\langle \left\{ \delta J(t), \delta J(0) \right\} \right\rangle \;, \label{eq:current-noise}
\end{eqnarray}
where the fluctuation operator 
for a quantity $A$ 
is given by $\delta A \equiv A - \langle A \rangle$,
and $\{ A, B \} \equiv AB +BA$ is anticommutator.
Conventionally, the shot noise has been defined 
by the value of $S$ at zero temperature $T=0$, 
where the thermal noise is completely suppressed.
This definition of the shot noise has been successfully exploited 
for studying the properties at low temperatures
\cite{PhysRep.336.1}.
It is difficult, however,
for interacting electron systems 
to separate the shot noise and thermal noise 
at finite temperatures.

On the basis of a generalized Kubo formalism,
Fujii has shown that the shot noise can be related
to a correlation function 
between the current fluctuation and the charge fluctuation,
\begin{eqnarray}
S_h \equiv - \left\langle \left\{ \delta J, e(\delta N_L - \delta N_R) \right\} \right\rangle \;, \label{eq:newshotnoise}
\end{eqnarray}
and has suggested this correlation function as a finite-temperature shot noise.
\cite{JPhysSocJpn.76.044709}.
The correlation function $S_h$ has been shown to satisfy the identity,
\begin{eqnarray}
S_{h} = S - 4 k_B T \, G \;, \label{eq:nonequilibriumKubo}
\end{eqnarray}
with the differential conductance $G$ 
and $S$ defined in  Eq.\ \eqref{eq:current-noise}.
The expression \eqref{eq:nonequilibriumKubo} 
is identical to an empirical formula that has been applied as an estimation of
the effective shot noise at finite temperatures
from the measurement of the current noise and the differential conductance.

It has also been confirmed that 
a number of the properties of the shot noise can be rederived
from the expression of $S_h$ in Eq.\ (\ref{eq:newshotnoise}).
First, at zero temperature, Eq.\ (\ref{eq:nonequilibriumKubo}) is 
simplified as $S_{h} = S$.
Therefore, $S_h$ defined in Eq.\ (\ref{eq:newshotnoise}) 
clearly agrees with the conventional shot noise defined at $T=0$ by $S$.
Second, it has been proved that
the shot noise defined by Eq.\ (\ref{eq:newshotnoise}) 
vanishes in the linear response regime ($V=0$): $S_h=0$ .
This means that the Nyquist-Johnson relation is also reproduced.
These observations show that 
Eq.\ (\ref{eq:newshotnoise}) can be regarded
as an extension of the shot noise
to all temperatures.

Therefore $S_h$ given in Eq. \eqref{eq:newshotnoise}
enables one to study finite-temperature effects on the shot noise directly, without calculating $S$ and $G$ separately.
In the present paper, however,
we focus on the shot noise at zero temperature. 
Specifically, we calculate $S_h$ exactly up to
$V^3$
in the particle-hole symmetric case where 
the occupation number of each orbital 
is given by $n_{dm}^{}=1/2$.
We demonstrate two different strategies to achieve our goal.
The first
starts from $S_h$ defined in
Eq.\ (\ref{eq:newshotnoise}) 
and reaches the result given in  Eq.\ \eqref{eq:shotnoise_fromC}.
The second
from Eq. \eqref{eq:current-noise}
yields the final expression given in Eq.\ \eqref{eq:shotnoise_fromC2}. 
The results obtained in these two ways,
Eq.\ \eqref{eq:shotnoise_fromC} and Eq.\ \eqref{eq:shotnoise_fromC2}, 
agree with each other naturally, although 
the contributions of each Feynman diagram for $S_h$ and 
that for $S$ do not have one-to-one correspondence 
as summarized in Table \ref{tab:S_part}.

\subsubsection{Shot noise $S_h$ at absolute zero}
\label{subsec:shot_noise_I}

We now calculate the shot noise from the expression given 
in Eq.\ (\ref{eq:newshotnoise}) and derive the leading asymptotic 
dependence of applied bias voltage at $T=0$
in the particle-hole symmetric case.

First, substituting the current operator Eq.\ (\ref{eq:current-operator}) and charge fluctuation operator
into Eq.\ \eqref{eq:newshotnoise},
the shot noise is readily expanded in the Keldysh formalism as,
\begin{eqnarray}
S_h = \left[ ( F_{hLL} - F_{hLR} ) + (L \leftrightarrow R) \right] +(\mbox{c.c.}) \;,
\end{eqnarray}
with,
\begin{widetext}
\begin{eqnarray}
F_{h\alpha \alpha'}
&=& -i \frac{e^2}{\hbar} \sum_{k,k',m,m'} V_{k\alpha m} \left\langle T_c \, {\cal S}_c \, c_{k \alpha m}^{\dagger}(0^+) d_{m}(0^+) c_{k' \alpha' m'}^{\dagger}(0^-)c_{k' \alpha' m'}(0^-)  \right\rangle_{{\rm connected}}^{} \;. \label{eq:partofS_h}
\end{eqnarray}
\end{widetext}
Here,
${\cal S}_c =  T_c \exp [-i \int_c dt\, {\cal H}_1^{} (t)]$ is
the time-evolution operator,
the Keldysh contour $c$ runs along the forward time direction
on the branch ``$-$" followed by the backward evolution on the branch ``$+$",
$T_c$ is the corresponding contour-ordering operator,
and
$\langle \dots \rangle_{{\rm connected}}^{}$
takes the sum of all connected diagrams.
The equal time correlation is defined by the Keldysh contour:
$0^{\pm}$ in Eq. \eqref{eq:partofS_h} is on the branch `$\pm$'.
The form of Eq.\ \eqref{eq:partofS_h} enables us
to apply perturbation expansion in $U$.
Then, the asymptotic behavior of the shot noise up to $V^3$ can be calculated 
using the RPT.

\begin{figure}[tb]
\includegraphics[width=7cm]{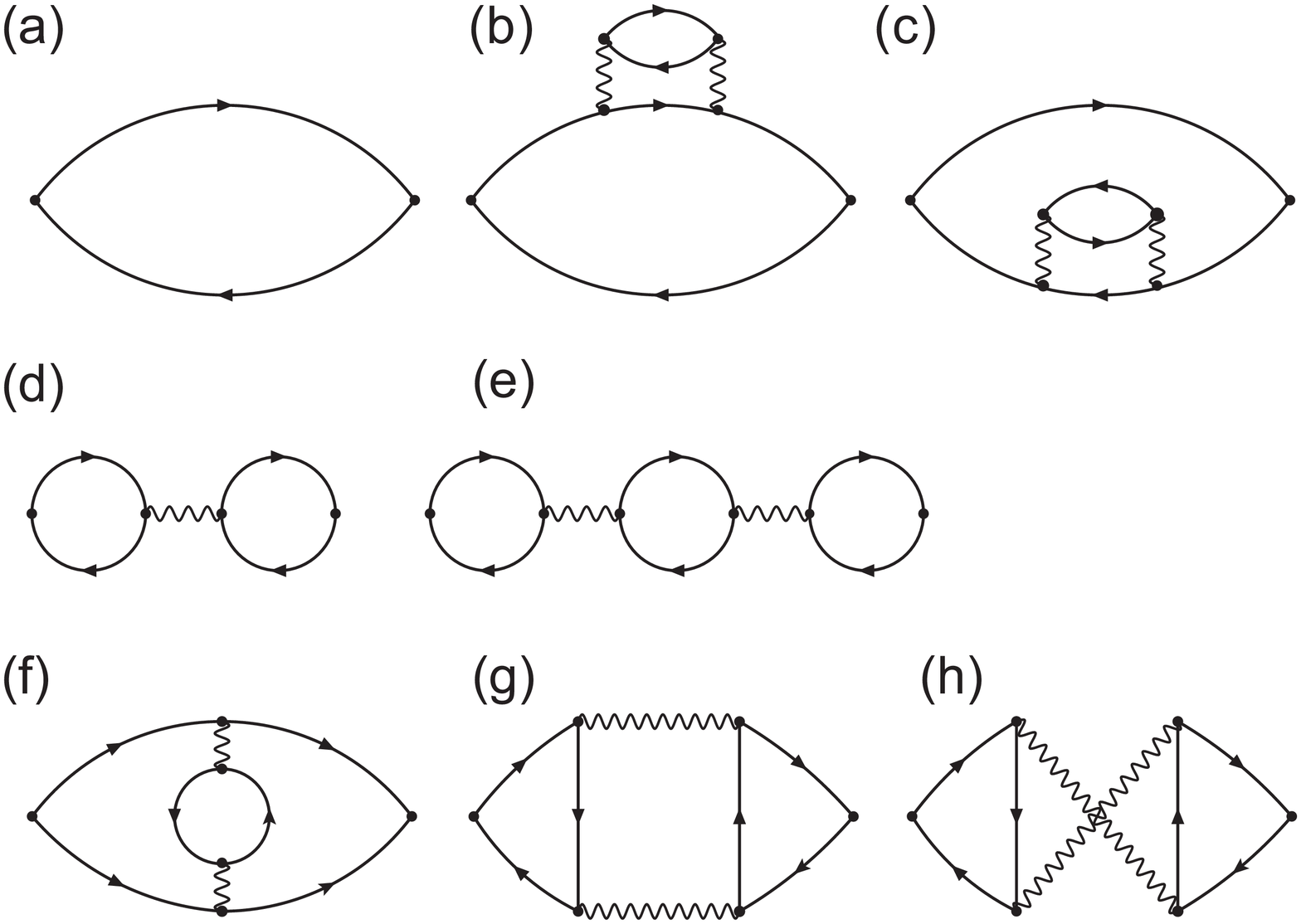}
\caption{\label{fig:diagrams}
The Feynman diagrams for $S_h$ and $S$ generated from
 Eqs. \eqref{eq:partofS_h} and \eqref{eq:currentnoise_parts}.
The contributions up to second order in $\bar{U}$ are shown.
The Hartree term is included in the nonperturbative Green's function.
Note that the fluctuation operators of the two verticies
at the left and right ends
for $S_h$ and those for $S$ are different.
}
\end{figure}

The contributions of the second order perturbation in $\bar{U}$ 
can be classified using the diagrams shown in Fig. \ref{fig:diagrams}.
Note that for $S_h$ defined in Eq.\ \eqref{eq:newshotnoise}
the fluctuation operator assigned for the left end of each diagram 
and that for the right end are different, i.e., 
one of the two is the current operator and the other 
is the charge fluctuation operator. 
The calculated contributions of each diagram are summarized in Table \ref{tab:diagram}.
In the particle-hole symmetry case 
the contributions from diagram (d) and (e)  vanish.
The contribution from diagram (g) and that 
from (f)  are the same 
for arbitrary parameters.
Furthermore 
the contribution from (h) coincides with that of (f)  and (g) 
 in the particle-hole symmetric case for the symmetric lead-dot coupling.
\begin{table*}
\caption{\label{tab:diagram}
The contributions of each diagram shown in Fig.\ \ref{fig:diagrams}
for $S_h$ and $S$.
The unit is $C=(2e^3/h)|V|(eV/\bar{\Gamma})^2$.
}
\begin{ruledtabular}
\begin{tabular}{cccccc}
diagram&(a)&(b)&(c)&(d),(e)&(f),(g),(h)\\
\hline
 $S_h$ & $N C/12$ & $N(N-1)(R-1)^2 C/24$ & $5N(N-1)(R-1)^2 C/24$ & 0& $N(N-1)(R-1)^2 C/6$ \\
\\
 $S$ & $N C/12$ & $N(N-1)(R-1)^2 C/8$ & $N(N-1)(R-1)^2 C/8$ & 0 & $N(N-1)(R-1)^2 C/6$ \\
\end{tabular}
\label{tab:S_part}
\end{ruledtabular}
\end{table*}

Collecting all these contributions, we obtain the total shot noise 
at $T=0$ up to order $V^3$,
\begin{eqnarray}
S_h &=& S_h^{(a)} + S_h^{(b)} + S_h^{(c)} + 3S_h^{(f)} \nonumber \\
&=& \frac{2Ne^3}{h}|V| \left( \frac{eV}{\bar{\Gamma}} \right)^2 \left[ \frac{1}{12} +\frac{3}{4}(N-1)(R-1)^2 \right] . \nonumber \\
\label{eq:shotnoise_fromC} 
\end{eqnarray}
For $N=2$, this result agrees with the one for the SU(2) Anderson model
\cite{PhysRevB.80.233103,JPhysSocJpn.79.044714}.
In the Kondo limit where $R\to N/(N-1)$, our result agrees with 
 Mora {\it et al.}'s result 
in the particle-hole symmetric case 
for the symmetric lead-dot coupling 
\cite{PhysRevB.80.155322}.

\subsubsection{Current noise $S$ at absolute zero}
\label{subsec:shot_noise_II}

Here we calculate the shot noise via the current noise $S$ defined in 
 Eq.\ \eqref{eq:current-noise}.
Because the thermal noise is completely suppressed at zero temperature, 
we can extract the pure shot noise from the current noise $S$.
We derive the asymptotic form of the shot noise 
along a similar line, as that described above.

Substituting the current operator Eq.\ \eqref{eq:current-operator} into Eq.\ \eqref{eq:current-noise} readily leads to the current noise expression in the Keldysh formalism as,
\begin{eqnarray}
S &=& \frac{2e^2}{\hbar^2} \int dt \left[ \left( F_{LL}^{+-} (t) - F_{LR}^{+-} (t) \right) + (L \leftrightarrow R) \right] \nonumber \\
&& \qquad\qquad\qquad\qquad\qquad\qquad + (\mbox{c.c.}) \;,
\end{eqnarray}
with,
\begin{widetext}
\begin{eqnarray}
F_{\alpha\alpha'}(t, t')
&=& i^2 \sum_{kk'mm'} \left[ V_{\alpha k}^{} V_{\alpha'k'}^* \left\langle T_c \, {\cal S}_c \,
c_{\alpha km}^{\dagger} (t) d_{m}^{} (t) c_{\alpha'k'm'}^{\dagger}(t') d_{m}^{} (t') \right\rangle_{{\rm connected}}^{} \right. \nonumber \\ 
&& \qquad \qquad \quad \left. - V_{\alpha k}^{} V_{\alpha'k'}^* \left\langle T_c \, \, {\cal S}_c \,
c_{\alpha km}^{\dagger} (t) d_{m}^{}(t) d_{m'}^{\dagger}(t') c_{\alpha'k'm'}^{} (t') \right\rangle_{{\rm connected}}^{} \right] \;.
\label{eq:currentnoise_parts}
\end{eqnarray}
\end{widetext}
The contributions of the second order perturbation in $\bar{U}$ 
can be classified using the diagrams shown in Fig.\ \ref{fig:diagrams}
for this correlation function.
Note that for the current noise $S$
both of the two verticies, at the left and right ends of each diagram, 
are given by the current fluctuation operator $\delta J$. 
The calculated contributions of each diagram are summarized
in Table \ref{tab:diagram}. 
For this reason, 
the contributions from the diagram (b) and those from (c) 
are the same for $S$, in contrast to the case for $S_h$.
Nevertheless, there are some similarities between the two cases. 
The contributions from (d) and (e) vanish
in the particle-hole symmetric case.
The contribution from diagram (g) and that from (f) are the same 
for arbitrary parameters, and the contribution from
(h) coincides with those of (g) or (f)
in the particle-hole symmetric case for the symmetric lead-dot coupling.
In the Kondo limit the contribution of each diagram 
agrees with that of the corresponding diagram used 
in Mora's calculations
\cite{PhysRevB.80.155322}.

Collecting all these contributions, 
we obtain the total current noise up to order $V^3$ as,
\begin{eqnarray}
S &=& S^{(a)} +2 S^{(b)} + 3 S^{(f)} \nonumber\\
&=&\frac{2Ne^3}{h}|V| \left( \frac{eV}{\bar{\Gamma}} \right)^2 \left[ \frac{1}{12} + \frac{3}{4}(N-1)(R-1)^2  \right] . \nonumber \\
\label{eq:shotnoise_fromC2}
\end{eqnarray}
This result agrees with
Eq.\ \eqref{eq:shotnoise_fromC} which has been deduced from $S_h$.

\subsection{Fano factor}\label{subsec:Fanofactor}

We consider the backscattering current $I_b$  which contains
all effects of the quantum and thermal fluctuations. 
It is defined via the deviation of the nonequilibrium current $I$ 
from the value in the unitary limit $Ne^2V/h$, as,
\begin{eqnarray}
I_b &\equiv& \frac{Ne^2}{h}V -I \nonumber \\
&=& \frac{Ne^2}{h}V \left( \frac{eV}{\bar{\Gamma}}\right)^2 \left[ \frac{1}{12}+ \frac{5(N-1)}{12} \left( R-1 \right)^2  \right] \;.
\end{eqnarray}
Here, we have used the result for $I$ given in Eq.\ \eqref{eq:current}.
Generally, the current noise for the forward current and that
for the backscattering current are equivalent in the case where the
systems is coupled to two leads.
Therefore, the Fano factor of the backscattering current 
can be expressed in the form,
\begin{eqnarray}
F_b \equiv \frac{S}{2eI_b} 
\, = \frac{1 + 9(N-1)(R-1)^2}{1 + 5(N-1)(R-1)^2}\;, \label{eq:fano}
\end{eqnarray}
and this is one of the main results of  the present work.

In the noninteracting case ($U= 0$), the Wilson ratio takes the 
value $R=1$, which leads to the Poisson noise of $S = 2eI_b$ and  $F_b=1$. 
For small $U$,
substituting the perturbation expansion for the Wilson ratio
given in Eq.\ \eqref{eq:WR-ex} into Eq.\ \eqref{eq:fano},
the Fano factor can be calculated exactly
up to terms of order $U^5$,
\begin{widetext}
\begin{eqnarray}
F_b &=&
1
+4 (N-1) u^2
-8 (N-1) (N-2) u^3
+2 (N-1) \biggl[ 6 N^2-\left(26+\pi ^2\right) N-3 \pi ^2+66 \biggr] u^4 \nonumber \\
&& \quad - (N-1) (N-2) \left[ 16 N^2 +  \left(84 \zeta (3) - 10 \pi^2- 160 \right) N + 616 \zeta (3) - \frac{178}{3} \pi ^2 +8 \right] u^5
+\mathcal{O} (u^6) \label{eq:FF-ex}
\;.
\end{eqnarray}
\end{widetext}
This expression obviously shows that
a larger orbital degeneracy makes the initial rise of $F_b$ steeper,  
as the coefficient of the order $u^2$ term increases with $N$.

\begin{table}
\caption{\label{tab:fano}
The Fano factor $F_b$ in the Kondo limit
($U\to\infty$),
for several choices of orbital degeneracy $N$.}
\begin{ruledtabular}
\begin{tabular}{cccccc}
$N$ & 2 & 4 & 6 & 8 & $\to \infty$ \\
\hline
$F_b$ & 5/3 & 3/2 & 7/5 & 4/3 & $\to 1$ \\
\end{tabular}
\end{ruledtabular}
\end{table}

In the opposite limit, namely, the Kondo limit ($U \to \infty$),
the Wilson ratio approaches to the universal value $R \to N/(N-1)$, 
and the Fano factor takes the form,
\begin{eqnarray}
F_b  \to \frac{N+8}{N+4} \label{eq:FanoKondoLimit}\;.
\end{eqnarray}
The explicit values for several $N$ are given in the Table \ref{tab:fano}.
Specifically in the limit of large orbital degeneracy $N \to \infty$, 
the shot noises take the Poisson value $S \to 2eI_b$,
and the Fano factor approaches to the noninteracting value 
$F_b \to 1$ even though the Coulomb repulsion 
has been taken first to be $U \to \infty$.
This originates from the fact that renormalization is weakened
in the limit of large $N$.
In practice,
renormalized parameters takes $\bar{U} \to 0$ and $\bar{\Gamma} \to \Gamma$
in the large $N$ limit ($R \to 1$)
of Eqs. \eqref{eq:Uren2} and \eqref{eq:Gam_ren2}.
For $N=2$, i.e., in the SU(2) case, our result is consistent with
the previous results obtained by Gogolin and Komnik, Sela {\it et al.} and Fujii.
\cite{PhysRevLett.97.016602,PhysRevLett.97.086601,JPhysSocJpn.79.044714}.

\begin{figure}[bt]
\includegraphics[width=6.4cm]{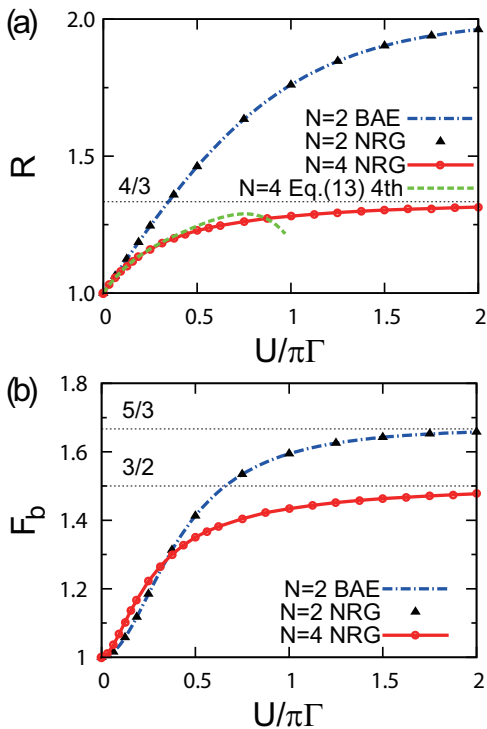}
\caption{\label{fig:fanofactor}
(Color online)
(a) The Wilson ratio $R$, and (b) the Fano factor $F_b$
as functions of the on-site Coulomb repulsion $U$.
The dashed-dotted line denotes the results deduced from the BAE for $N=2$.
The NRG results are plotted for $N=2$ ($\blacktriangle$),
and for $N=4$ with the solid circle ($\bullet$).
The dashed line in (a) is the Wilson ratio for $N=4$,
obtained from the perturbative expansion up to terms of order $U^4$ 
given in Eq.\ \eqref{eq:WR-ex}.
}
\end{figure}

In order to clarify the behavior of $F_b$ in the intermediate values of $U$,
we have carried out the NRG calculations for $N=2$ and $4$.
Specifically, we have deduced the Wilson ratio $R$ from 
the low-energy NRG fixed point
\cite{springerlink:10.1140/epjb/e2004-00256-0}.  
This method has also been applied to the two-channel Anderson model 
in the recent work of Nishikawa {\it el al.}
\cite{PhysRevB.82.115123}.
In the case of $N=2$, the NRG and Bethe ansatz results 
for the renormalized parameters have been shown to agree well
\cite{Wiegmann1980163,Kawakami1981483,PhysRevB.28.6904}
as seen in Fig.\ \ref{fig:fanofactor} (a).
There is no Bethe ansatz solutions for $N>2$
in the particle-hole symmetric case, 
still the NRG is applicable for $N=4$. 
We can see, in Fig.\ \ref{fig:fanofactor} (a), that 
the NRG results for the Wilson ratio
for $N=4$ are in good agreement with the perturbation result
up to a term of order $U^4$ given in Eq.\ \eqref{eq:WR-ex}
for a small Coulomb repulsion $u \lesssim 0.3$.
Furthermore, the NRG results approach the correct
universal value $R\to 4/3$ for larger $U$.
Therefore, the $U$ dependence of
the Fano factor for $N=4$ can also be deduced,
using expression \eqref{eq:fano}
with the nonperturbative NRG approach,
which is discussed below.

As the Coulomb repulsion increases further, 
the charge fluctuation in the dot is suppressed 
and the Wilson ratio increases rapidly 
from the noninteracting value $R=1$ to 
the universal value in the Kondo limit.
This crossover from the weak-coupling regime to the Kondo regime
is also observed in the Fano factor, shown 
in Fig.\ \ref{fig:fanofactor} (b),
as a rapid convergence of the Fano factor 
to the universal value
$F_b=5/3$ and $3/2$ for $N=2$ and $4$, respectively.
Note that the Wilson ratio is a decreasing function 
of $\chi_{c,d}^{\ast}/\chi_{d}^{\ast}$, namely, 
the ratio of the charge susceptibility to 
the susceptibility as shown in Eq.\ \eqref{eq:WR}.
Therefore, the charge fluctuation at the dot-site suppresses 
the Fano factor for small Coulomb repulsion $U \lesssim \pi \Gamma$.
The NRG results for $N=4$ have successfully revealed 
the precise feature of the crossover for the system 
with the orbital degeneracy.

\begin{figure}[bt]
\includegraphics[width=6.4cm]{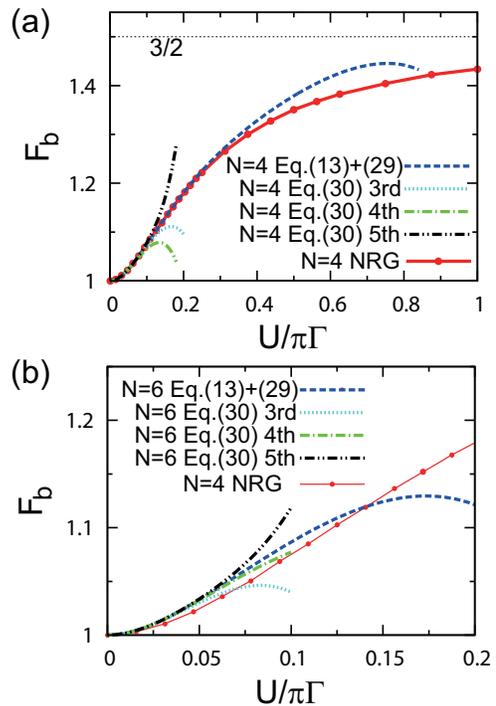}
\caption{\label{fig:FfRT}
(Color online) Perturbation results for the Fano factor 
for (a) $N=4$, and (b) $N=6$.
The dashed line is obtained just 
 by substituting Eq.\ \eqref{eq:WR-ex} into Eq.\ \eqref{eq:fano}.
The dotted, dashed-dotted, and dashed-dot-dotted lines 
denote the values from the series expansion 
up to terms of order $U^3$, $U^4$ and $U^5$, respectively, 
given in Eq.\ \eqref{eq:FF-ex}. 
The solid line with the solid circle ($\bullet$) is the NRG results for $N=4$.
}
\end{figure}

The perturbation series is applicable also for large orbital degeneracies $N$ 
where the NRG is not feasible. 
Therefore, it is worthwhile to discuss
the convergence of 
the series expansion.
For this purpose,
the Fano factor for $N=4$ is plotted in Fig. \ref{fig:FfRT} (a),
keeping the first few terms of the series given in Eq.\ \eqref{eq:FF-ex}.
We can see that the values of $F_b$ deduced from Eq.\ \eqref{eq:FF-ex}  
agree with the NRG result for a relatively narrow range 
of the Coulomb interaction $u \lesssim 0.1$, 
while the series expansion for the Wilson ratio 
quantitatively works in a wider range $u \lesssim 0.3$.
This discrepancy is caused by 
the expansion of the denominator on the right-hand side 
of Eq.\ \eqref{eq:fano} with respect to $(R-1)$, 
the convergence radius for which becomes small as shown in the Appendix.
This can be resolved, however, using 
the expression of the Fano factor Eq.\ \eqref{eq:fano}, 
as it is, without expanding the denominator,
and then substituting there 
the series expansion for the Wilson ration given 
in Eq.\ \eqref{eq:WR-ex}.
The results obtained in this way agree with 
the NRG data for $u \lesssim 0.3$,
as shown by the dashed line in Fig.\ \ref{fig:FfRT} (a).
Specifically, for $N=4$, both the coefficient of the order $u^3$ term
and that for the order $u^4$ term in Eq. \eqref{eq:FF-ex} become negative,
while that for the order $u^5$ term is positive.
For this reason, in Fig.\ \ref{fig:FfRT} (a),
the fourth order curve deviates from
the NRG result earlier than the third order curve,
and then the fifth order curve again approaches
the NRG result.
Finally, we examine the $N=6$ case,
for which nonperturbative approaches are not available at present.
The value of the Fano factor for $N=6$ deduced from 
the perturbation series are shown in Fig.\ \ref{fig:FfRT} (b).
We see that the results are quantitatively valid 
for $u \lesssim 0.1$,
although the convergence of 
the series expansion for $R$ 
and that for $F_b$ given in Eqs.\ \eqref{eq:WR-ex} 
and \eqref{eq:FF-ex} 
become worse with an increase of $N$. 
Nevertheless, the results show clearly that the initial rise of $F_b$ near 
$u\simeq 0.1$ is steeper for $N=6$ than that for $N=4$.

\section{discussion and summary}\label{sec:summary}

At the end of the previous section,
we have calculated the value of the Fano factor 
from the data for the Wilson ratio.
Conversely, it may be possible to determine
the value of the Wilson ratio from the measurements of the Fano factor,
using the expression,
\begin{eqnarray}
R = 1 + \sqrt{\frac{F_b-1}{(N-1)(9-5F_b)}} \;, \label{eq:WRinFF}
\end{eqnarray}
which is readily obtained by inverting Eq.\ (\ref{eq:fano}).
This scheme for estimating $R$ from $F_b$
does not need the value of the $g$-factor for the electrons in the dot.
Only the observation of $F_b$ determines the accuracy of observed $R$.
Alternatively, one 
can estimate the Wilson ratio
from the measurements of the susceptibilities 
$\chi_{c,d}^{\ast}/\chi_{d}^{\ast}$.
In this case, the $g$ factor is also needed
to determine them accurately.
Therefore, Eq. \eqref{eq:WRinFF} may enable us to experimentally observe the Wilson ratio with a more reasonable accuracy.

Finally, we comment on a physical interpretation for the values of 
the Fano factor for the orbital Kondo systems.
The expression for the Fano factor presented in Eq. \eqref{eq:fano} 
shows that effects of the Coulomb repulsion enter through the Wilson ratio $R$,
which characterizes the low-energy Fermi liquid state. 
The terms having $R$ in the coefficient come from
the vertex corrections and self-energy appearing in the RPT
approach to
the shot noise defined by Eqs. \eqref{eq:current-noise} or \eqref{eq:newshotnoise},
and the nonequilibrium current $I$.
In this aspect, the role of the Fano factor for the shot noise
resembles the role of the Stoner factor for the magnetic susceptibility.
These factors quantitatively enhance the response of the system 
against external fields, but do not change the qualitative feature 
of the low-energy states.
Therefore, the deviations of the Fano factor from the noninteracting value 
are caused by the scattering between the renormalized quasiparticles 
of the Fermi liquid, rather than 
an effective charge
as with other kinds of elementary excitations.

In summary, we have studied the current and shot noise
through the single quantum dot described by the multiorbital
impurity Anderson model, 
using
the nonequilibrium Kubo formalism.
Employing the renormalized perturbation theory, 
the current and the shot noise have been calculated
exactly in the particle-hole symmetric case 
up to order $V^3$ at $T=0$. 
The result for the shot noise
is given by Eq. \eqref{eq:shotnoise_fromC}, 
and then the  Fano factor $F_b$ of the backscattering current 
is determined by two parameters, i.e., the Wilson ratio $R$ and  
the orbital degeneracy $N$, as shown in Eq.\ \eqref{eq:fano}. 
We have also presented the explicit form of the Fano factor 
for small Coulomb repulsions up to terms of order $U^5$.
Furthermore we have deduced the dependence of $F_b$ for $N=2$ and 
$N=4$  on  the Coulomb repulsion,
using the Wilson ratio obtained with the NRG approach.
The Fano factor varies monotonically with an increase of $U$ 
from the noninteracting value 
$F_b=1$ until almost saturating for $U \gtrsim \pi \Gamma$ to the value 
in the Kondo limit,
$F_b=(N+8)/(N+4)$. 
For small $U$, the charge fluctuation at the dot-site suppresses 
the Fano factor.

\begin{acknowledgments}
The authors thank N. Kawakami, K. Kobayashi, Y. Nishikawa, 
A. C. Hewson and M. Tsuji for fruitful discussion,
and R. S. Deacon for reading the entire text in its original form.
One of us (RS) thanks S. Tarucha for warm encouragement
and was supported by
a Grant-in-Aid for JSPS Fellows.
One of us (AO) thanks the JSPS Grant-in-Aid 
for Scientific Research (C) support  (Grant No.\ 20540319).
Numerical computation was partly carried out
at the Yukawa Institute Computer Facility.
\end{acknowledgments}

\appendix

\section{
The convergence of the Fano factor \eqref{eq:FF-ex}
}

We discuss the
convergence of
the Fano factor as a power series of $U$ given in \eqref{eq:FF-ex}.

First, the Fano factor given in \eqref{eq:fano} can be expanded as a power series in $R-1$ around $R-1=0$, as,
\begin{eqnarray}
F_b = \left[ 1+9(N-1)(R-1)^2 \right] \sum_{n=0}^{\infty} C_n(N) (R-1)^{2n} \;,
\label{eq:FFinR}
\end{eqnarray}
with $C_n(N) = \left[ -5 \left( N-1 \right) \right]^n$.
and the radius of convergence,
\begin{eqnarray}
|R-1| < \sqrt{\frac{1}{5(N-1)}} \;. \label{eq:WRRC}
\end{eqnarray}
It follows from Eq. \eqref{eq:FFinR} that for large $N$ the power series slowly converges.
In addition, the Wilson ratio as a power series in $U$
given in Eq.\ \eqref{eq:WR-ex} slowly converges for large $N$.
Therefore, it can be concluded that
the convergence of the Fano factor as a power series in $U$
given in Eq.\ \eqref{eq:FF-ex} is slower for larger degeneracy $N$.

Second, we consider the radius of convergence of Eq.\ \eqref{eq:FF-ex}.
As mentioned in the text, the value of the Wilson ratio 
is bounded in a range,
\begin{eqnarray}
0 \le R-1 < \frac{1}{N-1} \;, \label{eq:WRrange}
\end{eqnarray}
for $0 \le U < \infty$.
For $N=2,4$,
it follows from Eqs. \eqref{eq:WRRC} and \eqref{eq:WRrange} that the radius of convergence of the Fano factor as a power series of $U$ is determined by,
\begin{eqnarray}
0 \le R-1 < \sqrt{\frac{1}{5(N-1)}} \;.
\end{eqnarray}
Therefore,
the radius of convergence can be obtained
as $u \sim 0.48$ for $N=2$ and $u \sim 0.72$ for $N=4$
from the Wilson ratio data shown in Fig.\ \ref{fig:fanofactor} (a).
However, for $N \geq 6$, the range of $R-1$ given in Eq. \eqref{eq:WRrange}
is narrower than that of Eq. \eqref{eq:WRRC}.
Therefore, the Fano factor for $N \geq 6$ can be written
as a power series of $U$ for arbitrarily positive $U$,
similarly to the universal quantities of the Kondo effect.

\newpage 

\begin{thebibliography}{63}
\expandafter\ifx\csname natexlab\endcsname\relax\def\natexlab#1{#1}\fi
\expandafter\ifx\csname bibnamefont\endcsname\relax
  \def\bibnamefont#1{#1}\fi
\expandafter\ifx\csname bibfnamefont\endcsname\relax
  \def\bibfnamefont#1{#1}\fi
\expandafter\ifx\csname citenamefont\endcsname\relax
  \def\citenamefont#1{#1}\fi
\expandafter\ifx\csname url\endcsname\relax
  \def\url#1{\texttt{#1}}\fi
\expandafter\ifx\csname urlprefix\endcsname\relax\def\urlprefix{URL }\fi
\providecommand{\bibinfo}[2]{#2}
\providecommand{\eprint}[2][]{\url{#2}}

\bibitem[{\citenamefont{Hewson}(1993{\natexlab{a}})}]{HewsonBook}
\bibinfo{author}{\bibfnamefont{A.~C.} \bibnamefont{Hewson}},
  \emph{\bibinfo{title}{The Kondo Problem to Heavy Fermions}}
  (\bibinfo{publisher}{Cambridge University Press},
  \bibinfo{year}{1993}{\natexlab{a}}).

\bibitem[{\citenamefont{Grobis et~al.}(2008)\citenamefont{Grobis, Rau, Potok,
  Shtrikman, and Goldhaber-Gordon}}]{PhysRevLett.100.246601}
\bibinfo{author}{\bibfnamefont{M.}~\bibnamefont{Grobis}},
  \bibinfo{author}{\bibfnamefont{I.~G.} \bibnamefont{Rau}},
  \bibinfo{author}{\bibfnamefont{R.~M.} \bibnamefont{Potok}},
  \bibinfo{author}{\bibfnamefont{H.}~\bibnamefont{Shtrikman}},
  \bibnamefont{and}
  \bibinfo{author}{\bibfnamefont{D.}~\bibnamefont{Goldhaber-Gordon}},
  \bibinfo{journal}{Phys. Rev. Lett.} \textbf{\bibinfo{volume}{100}},
  \bibinfo{pages}{246601} (\bibinfo{year}{2008}).

\bibitem[{\citenamefont{Scott et~al.}(2009)\citenamefont{Scott, Keane, Ciszek,
  Tour, and Natelson}}]{PhysRevB.79.165413}
\bibinfo{author}{\bibfnamefont{G.~D.} \bibnamefont{Scott}},
  \bibinfo{author}{\bibfnamefont{Z.~K.} \bibnamefont{Keane}},
  \bibinfo{author}{\bibfnamefont{J.~W.} \bibnamefont{Ciszek}},
  \bibinfo{author}{\bibfnamefont{J.~M.} \bibnamefont{Tour}}, \bibnamefont{and}
  \bibinfo{author}{\bibfnamefont{D.}~\bibnamefont{Natelson}},
  \bibinfo{journal}{Phys. Rev. B} \textbf{\bibinfo{volume}{79}},
  \bibinfo{pages}{165413} (\bibinfo{year}{2009}).

\bibitem[{\citenamefont{Schiller and Hershfield}(1998)}]{PhysRevB.58.14978}
\bibinfo{author}{\bibfnamefont{A.}~\bibnamefont{Schiller}} \bibnamefont{and}
  \bibinfo{author}{\bibfnamefont{S.}~\bibnamefont{Hershfield}},
  \bibinfo{journal}{Phys. Rev. B} \textbf{\bibinfo{volume}{58}},
  \bibinfo{pages}{14978} (\bibinfo{year}{1998}).

\bibitem[{\citenamefont{Meir and Golub}(2002)}]{PhysRevLett.88.116802}
\bibinfo{author}{\bibfnamefont{Y.}~\bibnamefont{Meir}} \bibnamefont{and}
  \bibinfo{author}{\bibfnamefont{A.}~\bibnamefont{Golub}},
  \bibinfo{journal}{Phys. Rev. Lett.} \textbf{\bibinfo{volume}{88}},
  \bibinfo{pages}{116802} (\bibinfo{year}{2002}).

\bibitem[{\citenamefont{Dong and Lei}(2002)}]{0953-8984-14-19-318}
\bibinfo{author}{\bibfnamefont{B.}~\bibnamefont{Dong}} \bibnamefont{and}
  \bibinfo{author}{\bibfnamefont{X.~L.} \bibnamefont{Lei}},
  \bibinfo{journal}{J. Phys: Condens. Matter} \textbf{\bibinfo{volume}{14}},
  \bibinfo{pages}{4963} (\bibinfo{year}{2002}).

\bibitem[{\citenamefont{L\'opez and S\'anchez}(2003)}]{PhysRevLett.90.116602}
\bibinfo{author}{\bibfnamefont{R.}~\bibnamefont{L\'opez}} \bibnamefont{and}
  \bibinfo{author}{\bibfnamefont{D.}~\bibnamefont{S\'anchez}},
  \bibinfo{journal}{Phys. Rev. Lett.} \textbf{\bibinfo{volume}{90}},
  \bibinfo{pages}{116602} (\bibinfo{year}{2003}).

\bibitem[{\citenamefont{Golub}(2006)}]{PhysRevB.73.233310}
\bibinfo{author}{\bibfnamefont{A.}~\bibnamefont{Golub}},
  \bibinfo{journal}{Phys. Rev. B} \textbf{\bibinfo{volume}{73}},
  \bibinfo{pages}{233310} (\bibinfo{year}{2006}).

\bibitem[{\citenamefont{Gogolin and
  Komnik}(2006{\natexlab{a}})}]{PhysRevB.73.195301}
\bibinfo{author}{\bibfnamefont{A.~O.} \bibnamefont{Gogolin}} \bibnamefont{and}
  \bibinfo{author}{\bibfnamefont{A.}~\bibnamefont{Komnik}},
  \bibinfo{journal}{Phys. Rev. B} \textbf{\bibinfo{volume}{73}},
  \bibinfo{pages}{195301} (\bibinfo{year}{2006}{\natexlab{a}}).

\bibitem[{\citenamefont{Moca et~al.}(2010)\citenamefont{Moca, Weymann, and
  Zar\'and}}]{PhysRevB.81.241305}
\bibinfo{author}{\bibfnamefont{C.~P.} \bibnamefont{Moca}},
  \bibinfo{author}{\bibfnamefont{I.}~\bibnamefont{Weymann}}, \bibnamefont{and}
  \bibinfo{author}{\bibfnamefont{G.}~\bibnamefont{Zar\'and}},
  \bibinfo{journal}{Phys. Rev. B} \textbf{\bibinfo{volume}{81}},
  \bibinfo{pages}{241305} (\bibinfo{year}{2010}).

\bibitem[{\citenamefont{de~Picciotto et~al.}(1997)\citenamefont{de~Picciotto,
  Reznikov, Heiblum, Umansky, Bunin, and Mahalu}}]{Nature.389.162}
\bibinfo{author}{\bibfnamefont{R.}~\bibnamefont{de~Picciotto}},
  \bibinfo{author}{\bibfnamefont{M.}~\bibnamefont{Reznikov}},
  \bibinfo{author}{\bibfnamefont{M.}~\bibnamefont{Heiblum}},
  \bibinfo{author}{\bibfnamefont{V.}~\bibnamefont{Umansky}},
  \bibinfo{author}{\bibfnamefont{G.}~\bibnamefont{Bunin}}, \bibnamefont{and}
  \bibinfo{author}{\bibfnamefont{D.}~\bibnamefont{Mahalu}},
  \bibinfo{journal}{Nature} \textbf{\bibinfo{volume}{389}},
  \bibinfo{pages}{162} (\bibinfo{year}{1997}).

\bibitem[{\citenamefont{Saminadayar et~al.}(1997)\citenamefont{Saminadayar,
  Glattli, Jin, and Etienne}}]{PhysRevLett.79.2526}
\bibinfo{author}{\bibfnamefont{L.}~\bibnamefont{Saminadayar}},
  \bibinfo{author}{\bibfnamefont{D.~C.} \bibnamefont{Glattli}},
  \bibinfo{author}{\bibfnamefont{Y.}~\bibnamefont{Jin}}, \bibnamefont{and}
  \bibinfo{author}{\bibfnamefont{B.}~\bibnamefont{Etienne}},
  \bibinfo{journal}{Phys. Rev. Lett.} \textbf{\bibinfo{volume}{79}},
  \bibinfo{pages}{2526} (\bibinfo{year}{1997}).

\bibitem[{\citenamefont{Lefloch et~al.}(2003)\citenamefont{Lefloch, Hoffmann,
  Sanquer, and Quirion}}]{PhysRevLett.90.067002}
\bibinfo{author}{\bibfnamefont{F.}~\bibnamefont{Lefloch}},
  \bibinfo{author}{\bibfnamefont{C.}~\bibnamefont{Hoffmann}},
  \bibinfo{author}{\bibfnamefont{M.}~\bibnamefont{Sanquer}}, \bibnamefont{and}
  \bibinfo{author}{\bibfnamefont{D.}~\bibnamefont{Quirion}},
  \bibinfo{journal}{Phys. Rev. Lett.} \textbf{\bibinfo{volume}{90}},
  \bibinfo{pages}{067002} (\bibinfo{year}{2003}).

\bibitem[{\citenamefont{Gogolin and
  Komnik}(2006{\natexlab{b}})}]{PhysRevLett.97.016602}
\bibinfo{author}{\bibfnamefont{A.~O.} \bibnamefont{Gogolin}} \bibnamefont{and}
  \bibinfo{author}{\bibfnamefont{A.}~\bibnamefont{Komnik}},
  \bibinfo{journal}{Phys. Rev. Lett.} \textbf{\bibinfo{volume}{97}},
  \bibinfo{pages}{016602} (\bibinfo{year}{2006}{\natexlab{b}}).

\bibitem[{\citenamefont{Sela et~al.}(2006)\citenamefont{Sela, Oreg, von Oppen,
  and Koch}}]{PhysRevLett.97.086601}
\bibinfo{author}{\bibfnamefont{E.}~\bibnamefont{Sela}},
  \bibinfo{author}{\bibfnamefont{Y.}~\bibnamefont{Oreg}},
  \bibinfo{author}{\bibfnamefont{F.}~\bibnamefont{von Oppen}},
  \bibnamefont{and} \bibinfo{author}{\bibfnamefont{J.}~\bibnamefont{Koch}},
  \bibinfo{journal}{Phys. Rev. Lett.} \textbf{\bibinfo{volume}{97}},
  \bibinfo{pages}{086601} (\bibinfo{year}{2006}).

\bibitem[{\citenamefont{Zarchin et~al.}(2008)\citenamefont{Zarchin, Zaffalon,
  Heiblum, Mahalu, and Umansky}}]{PhysRevB.77.241303}
\bibinfo{author}{\bibfnamefont{O.}~\bibnamefont{Zarchin}},
  \bibinfo{author}{\bibfnamefont{M.}~\bibnamefont{Zaffalon}},
  \bibinfo{author}{\bibfnamefont{M.}~\bibnamefont{Heiblum}},
  \bibinfo{author}{\bibfnamefont{D.}~\bibnamefont{Mahalu}}, \bibnamefont{and}
  \bibinfo{author}{\bibfnamefont{V.}~\bibnamefont{Umansky}},
  \bibinfo{journal}{Phys. Rev. B} \textbf{\bibinfo{volume}{77}},
  \bibinfo{pages}{241303} (\bibinfo{year}{2008}).

\bibitem[{\citenamefont{Yamauchi and Kobayashi}()}]{PCkobayashi}
\bibinfo{author}{\bibfnamefont{Y.}~\bibnamefont{Yamauchi}} \bibnamefont{and}
  \bibinfo{author}{\bibfnamefont{K.}~\bibnamefont{Kobayashi}},
  \bibinfo{note}{private communication}.

\bibitem[{\citenamefont{Sasaki et~al.}(2004)\citenamefont{Sasaki, Amaha,
  Asakawa, Eto, and Tarucha}}]{PhysRevLett.93.017205}
\bibinfo{author}{\bibfnamefont{S.}~\bibnamefont{Sasaki}},
  \bibinfo{author}{\bibfnamefont{S.}~\bibnamefont{Amaha}},
  \bibinfo{author}{\bibfnamefont{N.}~\bibnamefont{Asakawa}},
  \bibinfo{author}{\bibfnamefont{M.}~\bibnamefont{Eto}}, \bibnamefont{and}
  \bibinfo{author}{\bibfnamefont{S.}~\bibnamefont{Tarucha}},
  \bibinfo{journal}{Phys. Rev. Lett.} \textbf{\bibinfo{volume}{93}},
  \bibinfo{pages}{017205} (\bibinfo{year}{2004}).

\bibitem[{\citenamefont{Jarillo-Herrero
  et~al.}(2005)\citenamefont{Jarillo-Herrero, Kong, van~der Zant, Dekker,
  Kouwenhoven, and Francesch}}]{Nature434.484}
\bibinfo{author}{\bibfnamefont{P.}~\bibnamefont{Jarillo-Herrero}},
  \bibinfo{author}{\bibfnamefont{J.}~\bibnamefont{Kong}},
  \bibinfo{author}{\bibfnamefont{H.~S.~J.} \bibnamefont{van~der Zant}},
  \bibinfo{author}{\bibfnamefont{C.}~\bibnamefont{Dekker}},
  \bibinfo{author}{\bibfnamefont{L.~P.} \bibnamefont{Kouwenhoven}},
  \bibnamefont{and} \bibinfo{author}{\bibfnamefont{S.~D.}
  \bibnamefont{Francesch}}, \bibinfo{journal}{Nature}
  \textbf{\bibinfo{volume}{434}}, \bibinfo{pages}{484} (\bibinfo{year}{2005}).

\bibitem[{\citenamefont{Makarovski et~al.}(2007)\citenamefont{Makarovski,
  Zhukov, Liu, and Finkelstein}}]{PhysRevB.75.241407}
\bibinfo{author}{\bibfnamefont{A.}~\bibnamefont{Makarovski}},
  \bibinfo{author}{\bibfnamefont{A.}~\bibnamefont{Zhukov}},
  \bibinfo{author}{\bibfnamefont{J.}~\bibnamefont{Liu}}, \bibnamefont{and}
  \bibinfo{author}{\bibfnamefont{G.}~\bibnamefont{Finkelstein}},
  \bibinfo{journal}{Phys. Rev. B} \textbf{\bibinfo{volume}{75}},
  \bibinfo{pages}{241407} (\bibinfo{year}{2007}).

\bibitem[{\citenamefont{Izumida et~al.}(2001)\citenamefont{Izumida, Sakai, and
  Tarucha}}]{PhysRevLett.87.216803}
\bibinfo{author}{\bibfnamefont{W.}~\bibnamefont{Izumida}},
  \bibinfo{author}{\bibfnamefont{O.}~\bibnamefont{Sakai}}, \bibnamefont{and}
  \bibinfo{author}{\bibfnamefont{S.}~\bibnamefont{Tarucha}},
  \bibinfo{journal}{Phys. Rev. Lett.} \textbf{\bibinfo{volume}{87}},
  \bibinfo{pages}{216803} (\bibinfo{year}{2001}).

\bibitem[{\citenamefont{Borda et~al.}(2003)\citenamefont{Borda, Zar\'and,
  Hofstetter, Halperin, and von Delft}}]{PhysRevLett.90.026602}
\bibinfo{author}{\bibfnamefont{L.}~\bibnamefont{Borda}},
  \bibinfo{author}{\bibfnamefont{G.}~\bibnamefont{Zar\'and}},
  \bibinfo{author}{\bibfnamefont{W.}~\bibnamefont{Hofstetter}},
  \bibinfo{author}{\bibfnamefont{B.~I.} \bibnamefont{Halperin}},
  \bibnamefont{and} \bibinfo{author}{\bibfnamefont{J.}~\bibnamefont{von
  Delft}}, \bibinfo{journal}{Phys. Rev. Lett.} \textbf{\bibinfo{volume}{90}},
  \bibinfo{pages}{026602} (\bibinfo{year}{2003}).

\bibitem[{\citenamefont{Galpin et~al.}(2005)\citenamefont{Galpin, Logan, and
  Krishnamurthy}}]{PhysRevLett.94.186406}
\bibinfo{author}{\bibfnamefont{M.~R.} \bibnamefont{Galpin}},
  \bibinfo{author}{\bibfnamefont{D.~E.} \bibnamefont{Logan}}, \bibnamefont{and}
  \bibinfo{author}{\bibfnamefont{H.~R.} \bibnamefont{Krishnamurthy}},
  \bibinfo{journal}{Phys. Rev. Lett.} \textbf{\bibinfo{volume}{94}},
  \bibinfo{pages}{186406} (\bibinfo{year}{2005}).

\bibitem[{\citenamefont{Eto}(2005)}]{JPSJ.74.95}
\bibinfo{author}{\bibfnamefont{M.}~\bibnamefont{Eto}}, \bibinfo{journal}{J.
  Phys. Soc. Jpn.} \textbf{\bibinfo{volume}{74}}, \bibinfo{pages}{95}
  (\bibinfo{year}{2005}).

\bibitem[{\citenamefont{Choi et~al.}(2005)\citenamefont{Choi, L\'opez, and
  Aguado}}]{PhysRevLett.95.067204}
\bibinfo{author}{\bibfnamefont{M.-S.} \bibnamefont{Choi}},
  \bibinfo{author}{\bibfnamefont{R.}~\bibnamefont{L\'opez}}, \bibnamefont{and}
  \bibinfo{author}{\bibfnamefont{R.}~\bibnamefont{Aguado}},
  \bibinfo{journal}{Phys. Rev. Lett.} \textbf{\bibinfo{volume}{95}},
  \bibinfo{pages}{067204} (\bibinfo{year}{2005}).

\bibitem[{\citenamefont{Sakano and Kawakami}(2006)}]{PhysRevB.73.155332}
\bibinfo{author}{\bibfnamefont{R.}~\bibnamefont{Sakano}} \bibnamefont{and}
  \bibinfo{author}{\bibfnamefont{N.}~\bibnamefont{Kawakami}},
  \bibinfo{journal}{Phys. Rev. B} \textbf{\bibinfo{volume}{73}},
  \bibinfo{pages}{155332} (\bibinfo{year}{2006}).

\bibitem[{\citenamefont{Mravlje et~al.}(2006)\citenamefont{Mravlje,
  Ram\ifmmode~\check{s}\else \v{s}\fi{}ak, and Rejec}}]{PhysRevB.73.241305}
\bibinfo{author}{\bibfnamefont{J.}~\bibnamefont{Mravlje}},
  \bibinfo{author}{\bibfnamefont{A.}~\bibnamefont{Ram\ifmmode~\check{s}\else
  \v{s}\fi{}ak}}, \bibnamefont{and}
  \bibinfo{author}{\bibfnamefont{T.}~\bibnamefont{Rejec}},
  \bibinfo{journal}{Phys. Rev. B} \textbf{\bibinfo{volume}{73}},
  \bibinfo{pages}{241305} (\bibinfo{year}{2006}).

\bibitem[{\citenamefont{Lim et~al.}(2006)\citenamefont{Lim, Choi, Choi,
  L\'opez, and Aguado}}]{PhysRevB.74.205119}
\bibinfo{author}{\bibfnamefont{J.~S.} \bibnamefont{Lim}},
  \bibinfo{author}{\bibfnamefont{M.-S.} \bibnamefont{Choi}},
  \bibinfo{author}{\bibfnamefont{M.~Y.} \bibnamefont{Choi}},
  \bibinfo{author}{\bibfnamefont{R.}~\bibnamefont{L\'opez}}, \bibnamefont{and}
  \bibinfo{author}{\bibfnamefont{R.}~\bibnamefont{Aguado}},
  \bibinfo{journal}{Phys. Rev. B} \textbf{\bibinfo{volume}{74}},
  \bibinfo{pages}{205119} (\bibinfo{year}{2006}).

\bibitem[{\citenamefont{Kita et~al.}(2008)\citenamefont{Kita, Sakano, Ohashi,
  and Suga}}]{JPSJ.77.094707}
\bibinfo{author}{\bibfnamefont{T.}~\bibnamefont{Kita}},
  \bibinfo{author}{\bibfnamefont{R.}~\bibnamefont{Sakano}},
  \bibinfo{author}{\bibfnamefont{T.}~\bibnamefont{Ohashi}}, \bibnamefont{and}
  \bibinfo{author}{\bibfnamefont{S.}~\bibnamefont{Suga}}, \bibinfo{journal}{J.
  Phys. Soc. Jpn.} \textbf{\bibinfo{volume}{77}}, \bibinfo{pages}{094707}
  (\bibinfo{year}{2008}).

\bibitem[{\citenamefont{Numata et~al.}(2009)\citenamefont{Numata, Nisikawa,
  Oguri, and Hewson}}]{PhysRevB.80.155330}
\bibinfo{author}{\bibfnamefont{T.}~\bibnamefont{Numata}},
  \bibinfo{author}{\bibfnamefont{Y.}~\bibnamefont{Nisikawa}},
  \bibinfo{author}{\bibfnamefont{A.}~\bibnamefont{Oguri}}, \bibnamefont{and}
  \bibinfo{author}{\bibfnamefont{A.~C.} \bibnamefont{Hewson}},
  \bibinfo{journal}{Phys. Rev. B} \textbf{\bibinfo{volume}{80}},
  \bibinfo{pages}{155330} (\bibinfo{year}{2009}).

\bibitem[{\citenamefont{Yoshimori}(1976)}]{PTP.55.67}
\bibinfo{author}{\bibfnamefont{A.}~\bibnamefont{Yoshimori}},
  \bibinfo{journal}{Prog. Theor. Phys.} \textbf{\bibinfo{volume}{55}},
  \bibinfo{pages}{67} (\bibinfo{year}{1976}).

\bibitem[{\citenamefont{Vitushinsky et~al.}(2008)\citenamefont{Vitushinsky,
  Clerk, and Le~Hur}}]{PhysRevLett.100.036603}
\bibinfo{author}{\bibfnamefont{P.}~\bibnamefont{Vitushinsky}},
  \bibinfo{author}{\bibfnamefont{A.~A.} \bibnamefont{Clerk}}, \bibnamefont{and}
  \bibinfo{author}{\bibfnamefont{K.}~\bibnamefont{Le~Hur}},
  \bibinfo{journal}{Phys. Rev. Lett.} \textbf{\bibinfo{volume}{100}},
  \bibinfo{pages}{036603} (\bibinfo{year}{2008}).

\bibitem[{\citenamefont{Mora et~al.}(2008)\citenamefont{Mora, Leyronas, and
  Regnault}}]{PhysRevLett.100.036604}
\bibinfo{author}{\bibfnamefont{C.}~\bibnamefont{Mora}},
  \bibinfo{author}{\bibfnamefont{X.}~\bibnamefont{Leyronas}}, \bibnamefont{and}
  \bibinfo{author}{\bibfnamefont{N.}~\bibnamefont{Regnault}},
  \bibinfo{journal}{Phys. Rev. Lett.} \textbf{\bibinfo{volume}{100}},
  \bibinfo{pages}{036604} (\bibinfo{year}{2008}).

\bibitem[{\citenamefont{Mora}(2009)}]{PhysRevB.80.125304}
\bibinfo{author}{\bibfnamefont{C.}~\bibnamefont{Mora}}, \bibinfo{journal}{Phys.
  Rev. B} \textbf{\bibinfo{volume}{80}}, \bibinfo{pages}{125304}
  (\bibinfo{year}{2009}).

\bibitem[{\citenamefont{Mora et~al.}(2009)\citenamefont{Mora, Vitushinsky,
  Leyronas, Clerk, and Le~Hur}}]{PhysRevB.80.155322}
\bibinfo{author}{\bibfnamefont{C.}~\bibnamefont{Mora}},
  \bibinfo{author}{\bibfnamefont{P.}~\bibnamefont{Vitushinsky}},
  \bibinfo{author}{\bibfnamefont{X.}~\bibnamefont{Leyronas}},
  \bibinfo{author}{\bibfnamefont{A.~A.} \bibnamefont{Clerk}}, \bibnamefont{and}
  \bibinfo{author}{\bibfnamefont{K.}~\bibnamefont{Le~Hur}},
  \bibinfo{journal}{Phys. Rev. B} \textbf{\bibinfo{volume}{80}},
  \bibinfo{pages}{155322} (\bibinfo{year}{2009}).

\bibitem[{\citenamefont{Delattre et~al.}(2009)\citenamefont{Delattre,
  Feuillet-Palma, Herrmann, Morfin, Berroir, F\`{e}ve, Pla\c{c}ais, Glattli,
  Choi, Mora et~al.}}]{NaturePhys5.208}
\bibinfo{author}{\bibfnamefont{T.}~\bibnamefont{Delattre}},
  \bibinfo{author}{\bibfnamefont{C.}~\bibnamefont{Feuillet-Palma}},
  \bibinfo{author}{\bibfnamefont{L.~G.} \bibnamefont{Herrmann}},
  \bibinfo{author}{\bibfnamefont{P.}~\bibnamefont{Morfin}},
  \bibinfo{author}{\bibfnamefont{J.-M.} \bibnamefont{Berroir}},
  \bibinfo{author}{\bibfnamefont{G.}~\bibnamefont{F\`{e}ve}},
  \bibinfo{author}{\bibfnamefont{B.}~\bibnamefont{Pla\c{c}ais}},
  \bibinfo{author}{\bibfnamefont{D.~C.} \bibnamefont{Glattli}},
  \bibinfo{author}{\bibfnamefont{M.-S.} \bibnamefont{Choi}},
  \bibinfo{author}{\bibfnamefont{C.}~\bibnamefont{Mora}}, \bibnamefont{et~al.},
  \bibinfo{journal}{Nature Physics} \textbf{\bibinfo{volume}{5}},
  \bibinfo{pages}{208} (\bibinfo{year}{2009}).

\bibitem[{\citenamefont{Lipi\ifmmode~\acute{n}\else \'{n}\fi{}ski and
  Krychowski}(2010)}]{PhysRevB.81.115327}
\bibinfo{author}{\bibfnamefont{S.}~\bibnamefont{Lipi\ifmmode~\acute{n}\else
  \'{n}\fi{}ski}} \bibnamefont{and}
  \bibinfo{author}{\bibfnamefont{D.}~\bibnamefont{Krychowski}},
  \bibinfo{journal}{Phys. Rev. B} \textbf{\bibinfo{volume}{81}},
  \bibinfo{pages}{115327} (\bibinfo{year}{2010}).

\bibitem[{\citenamefont{Fujii}(2010)}]{JPhysSocJpn.79.044714}
\bibinfo{author}{\bibfnamefont{T.}~\bibnamefont{Fujii}}, \bibinfo{journal}{J.
  Phys. Soc. Jpn.} \textbf{\bibinfo{volume}{79}}, \bibinfo{pages}{044714}
  (\bibinfo{year}{2010}).

\bibitem[{\citenamefont{Hewson}(1993{\natexlab{b}})}]{PhysRevLett.70.4007}
\bibinfo{author}{\bibfnamefont{A.~C.} \bibnamefont{Hewson}},
  \bibinfo{journal}{Phys. Rev. Lett.} \textbf{\bibinfo{volume}{70}},
  \bibinfo{pages}{4007} (\bibinfo{year}{1993}{\natexlab{b}}).

\bibitem[{\citenamefont{Hewson}(1993{\natexlab{c}})}]{0953-8984-5-34-014}
\bibinfo{author}{\bibfnamefont{A.~C.} \bibnamefont{Hewson}},
  \bibinfo{journal}{J. Phys: Condens. Matter} \textbf{\bibinfo{volume}{5}},
  \bibinfo{pages}{6277} (\bibinfo{year}{1993}{\natexlab{c}}).

\bibitem[{\citenamefont{Oguri}(2001)}]{PhysRevB.64.153305}
\bibinfo{author}{\bibfnamefont{A.}~\bibnamefont{Oguri}},
  \bibinfo{journal}{Phys. Rev. B} \textbf{\bibinfo{volume}{64}},
  \bibinfo{pages}{153305} (\bibinfo{year}{2001}).

\bibitem[{\citenamefont{Hewson et~al.}(2005)\citenamefont{Hewson, Bauer, and
  Oguri}}]{JPhysCondMatt.17.5413}
\bibinfo{author}{\bibfnamefont{A.~C.} \bibnamefont{Hewson}},
  \bibinfo{author}{\bibfnamefont{J.}~\bibnamefont{Bauer}}, \bibnamefont{and}
  \bibinfo{author}{\bibfnamefont{A.}~\bibnamefont{Oguri}}, \bibinfo{journal}{J.
  Phys.: Condens. Matter} \textbf{\bibinfo{volume}{17}}, \bibinfo{pages}{5413}
  (\bibinfo{year}{2005}).

\bibitem[{\citenamefont{Fujii}(2007)}]{JPhysSocJpn.76.044709}
\bibinfo{author}{\bibfnamefont{T.}~\bibnamefont{Fujii}}, \bibinfo{journal}{J.
  Phys. Soc. Jpn.} \textbf{\bibinfo{volume}{76}}, \bibinfo{pages}{044714}
  (\bibinfo{year}{2007}).

\bibitem[{\citenamefont{Peskin and Schroeder}(1995)}]{FieldTheoryBook}
\bibinfo{author}{\bibfnamefont{M.~E.} \bibnamefont{Peskin}} \bibnamefont{and}
  \bibinfo{author}{\bibfnamefont{D.~V.} \bibnamefont{Schroeder}},
  \emph{\bibinfo{title}{An Introduction to Quantum Field Theory}}
  (\bibinfo{publisher}{Perseus, Reading Massachusetts}, \bibinfo{year}{1995}).

\bibitem[{\citenamefont{Oguri}(2005)}]{JPSJ.74.110}
\bibinfo{author}{\bibfnamefont{A.}~\bibnamefont{Oguri}}, \bibinfo{journal}{J.
  Phys. Soc. Jpn.} \textbf{\bibinfo{volume}{74}}, \bibinfo{pages}{110}
  (\bibinfo{year}{2005}).

\bibitem[{\citenamefont{Yosida and Yamada}(1970)}]{PTP46.244}
\bibinfo{author}{\bibfnamefont{K.}~\bibnamefont{Yosida}} \bibnamefont{and}
  \bibinfo{author}{\bibfnamefont{K.}~\bibnamefont{Yamada}},
  \bibinfo{journal}{Prog. Theor. Phys. Supplement}
  \textbf{\bibinfo{volume}{46}}, \bibinfo{pages}{244} (\bibinfo{year}{1970}).

\bibitem[{\citenamefont{Hershfield et~al.}(1991)\citenamefont{Hershfield,
  Davies, and Wilkins}}]{PhysRevLett.67.3720}
\bibinfo{author}{\bibfnamefont{S.}~\bibnamefont{Hershfield}},
  \bibinfo{author}{\bibfnamefont{J.~H.} \bibnamefont{Davies}},
  \bibnamefont{and} \bibinfo{author}{\bibfnamefont{J.~W.}
  \bibnamefont{Wilkins}}, \bibinfo{journal}{Phys. Rev. Lett.}
  \textbf{\bibinfo{volume}{67}}, \bibinfo{pages}{3720} (\bibinfo{year}{1991}).

\bibitem[{\citenamefont{Fujii and Ueda}(2003)}]{PhysRevB.68.155310}
\bibinfo{author}{\bibfnamefont{T.}~\bibnamefont{Fujii}} \bibnamefont{and}
  \bibinfo{author}{\bibfnamefont{K.}~\bibnamefont{Ueda}},
  \bibinfo{journal}{Phys. Rev. B} \textbf{\bibinfo{volume}{68}},
  \bibinfo{pages}{155310} (\bibinfo{year}{2003}).

\bibitem[{\citenamefont{Hewson}(2001)}]{0953-8984-13-44-314}
\bibinfo{author}{\bibfnamefont{A.~C.} \bibnamefont{Hewson}},
  \bibinfo{journal}{J. Phys.: Condens. Matter} \textbf{\bibinfo{volume}{13}},
  \bibinfo{pages}{10011} (\bibinfo{year}{2001}).

\bibitem[{\citenamefont{Wilson}(1975)}]{RevModPhys.47.773}
\bibinfo{author}{\bibfnamefont{K.~G.} \bibnamefont{Wilson}},
  \bibinfo{journal}{Rev. Mod. Phys.} \textbf{\bibinfo{volume}{47}},
  \bibinfo{pages}{773} (\bibinfo{year}{1975}).

\bibitem[{\citenamefont{Krishna-murthy
  et~al.}(1980)\citenamefont{Krishna-murthy, Wilkins, and
  Wilson}}]{PhysRevB.21.1003}
\bibinfo{author}{\bibfnamefont{H.~R.} \bibnamefont{Krishna-murthy}},
  \bibinfo{author}{\bibfnamefont{J.~W.} \bibnamefont{Wilkins}},
  \bibnamefont{and} \bibinfo{author}{\bibfnamefont{K.~G.}
  \bibnamefont{Wilson}}, \bibinfo{journal}{Phys. Rev. B}
  \textbf{\bibinfo{volume}{21}}, \bibinfo{pages}{1003} (\bibinfo{year}{1980}).

\bibitem[{\citenamefont{Wiegmann}(1980)}]{Wiegmann1980163}
\bibinfo{author}{\bibfnamefont{P.}~\bibnamefont{Wiegmann}},
  \bibinfo{journal}{Phys. Lett. A} \textbf{\bibinfo{volume}{80}},
  \bibinfo{pages}{163} (\bibinfo{year}{1980}).

\bibitem[{\citenamefont{Kawakami and Okiji}(1981)}]{Kawakami1981483}
\bibinfo{author}{\bibfnamefont{N.}~\bibnamefont{Kawakami}} \bibnamefont{and}
  \bibinfo{author}{\bibfnamefont{A.}~\bibnamefont{Okiji}},
  \bibinfo{journal}{Phys. Lett. A} \textbf{\bibinfo{volume}{86}},
  \bibinfo{pages}{483} (\bibinfo{year}{1981}).

\bibitem[{\citenamefont{Hershfield et~al.}(1992)\citenamefont{Hershfield,
  Davies, and Wilkins}}]{PhysRevB.46.7046}
\bibinfo{author}{\bibfnamefont{S.}~\bibnamefont{Hershfield}},
  \bibinfo{author}{\bibfnamefont{J.~H.} \bibnamefont{Davies}},
  \bibnamefont{and} \bibinfo{author}{\bibfnamefont{J.~W.}
  \bibnamefont{Wilkins}}, \bibinfo{journal}{Phys. Rev. B}
  \textbf{\bibinfo{volume}{46}}, \bibinfo{pages}{7046} (\bibinfo{year}{1992}).

\bibitem[{\citenamefont{Meir and Wingreen}(1992)}]{PhysRevLett.68.2512}
\bibinfo{author}{\bibfnamefont{Y.}~\bibnamefont{Meir}} \bibnamefont{and}
  \bibinfo{author}{\bibfnamefont{N.~S.} \bibnamefont{Wingreen}},
  \bibinfo{journal}{Phys. Rev. Lett.} \textbf{\bibinfo{volume}{68}},
  \bibinfo{pages}{2512} (\bibinfo{year}{1992}).

\bibitem[{\citenamefont{Wingreen and Meir}(1994)}]{PhysRevB.49.11040}
\bibinfo{author}{\bibfnamefont{N.~S.} \bibnamefont{Wingreen}} \bibnamefont{and}
  \bibinfo{author}{\bibfnamefont{Y.}~\bibnamefont{Meir}},
  \bibinfo{journal}{Phys. Rev. B} \textbf{\bibinfo{volume}{49}},
  \bibinfo{pages}{11040} (\bibinfo{year}{1994}).

\bibitem[{\citenamefont{Hershfield}(1993)}]{PhysRevLett.70.2134}
\bibinfo{author}{\bibfnamefont{S.}~\bibnamefont{Hershfield}},
  \bibinfo{journal}{Phys. Rev. Lett.} \textbf{\bibinfo{volume}{70}},
  \bibinfo{pages}{2134} (\bibinfo{year}{1993}).

\bibitem[{\citenamefont{Oguri}(2002)}]{JPSJ.71.2969}
\bibinfo{author}{\bibfnamefont{A.}~\bibnamefont{Oguri}}, \bibinfo{journal}{J.
  Phys. Soc. Jpn.} \textbf{\bibinfo{volume}{71}}, \bibinfo{pages}{2969}
  (\bibinfo{year}{2002}).

\bibitem[{\citenamefont{Blanter and B\"uttiker}(2000)}]{PhysRep.336.1}
\bibinfo{author}{\bibfnamefont{Y.~M.} \bibnamefont{Blanter}} \bibnamefont{and}
  \bibinfo{author}{\bibfnamefont{M.}~\bibnamefont{B\"uttiker}},
  \bibinfo{journal}{Phys. Rep.} \textbf{\bibinfo{volume}{336}},
  \bibinfo{pages}{1} (\bibinfo{year}{2000}).

\bibitem[{\citenamefont{Sela and Malecki}(2009)}]{PhysRevB.80.233103}
\bibinfo{author}{\bibfnamefont{E.}~\bibnamefont{Sela}} \bibnamefont{and}
  \bibinfo{author}{\bibfnamefont{J.}~\bibnamefont{Malecki}},
  \bibinfo{journal}{Phys. Rev. B} \textbf{\bibinfo{volume}{80}},
  \bibinfo{pages}{233103} (\bibinfo{year}{2009}).

\bibitem[{\citenamefont{Hewson et~al.}(2004)\citenamefont{Hewson, Oguri, and
  Meyer}}]{springerlink:10.1140/epjb/e2004-00256-0}
\bibinfo{author}{\bibfnamefont{A.~C.} \bibnamefont{Hewson}},
  \bibinfo{author}{\bibfnamefont{A.}~\bibnamefont{Oguri}}, \bibnamefont{and}
  \bibinfo{author}{\bibfnamefont{D.}~\bibnamefont{Meyer}},
  \bibinfo{journal}{Eur. Phys. J. B} \textbf{\bibinfo{volume}{40}},
  \bibinfo{pages}{177} (\bibinfo{year}{2004}).

\bibitem[{\citenamefont{Nishikawa et~al.}(2010)\citenamefont{Nishikawa, Crow,
  and Hewson}}]{PhysRevB.82.115123}
\bibinfo{author}{\bibfnamefont{Y.}~\bibnamefont{Nishikawa}},
  \bibinfo{author}{\bibfnamefont{D.~J.~G.} \bibnamefont{Crow}},
  \bibnamefont{and} \bibinfo{author}{\bibfnamefont{A.~C.}
  \bibnamefont{Hewson}}, \bibinfo{journal}{Phys. Rev. B}
  \textbf{\bibinfo{volume}{82}}, \bibinfo{pages}{115123}
  (\bibinfo{year}{2010}).

\bibitem[{\citenamefont{Zlati\ifmmode~\acute{c}\else \'{c}\fi{} and
  Horvati\ifmmode~\acute{c}\else \'{c}\fi{}}(1983)}]{PhysRevB.28.6904}
\bibinfo{author}{\bibfnamefont{V.}~\bibnamefont{Zlati\ifmmode~\acute{c}\else
  \'{c}\fi{}}} \bibnamefont{and}
  \bibinfo{author}{\bibfnamefont{B.}~\bibnamefont{Horvati\ifmmode~\acute{c}\else
  \'{c}\fi{}}}, \bibinfo{journal}{Phys. Rev. B} \textbf{\bibinfo{volume}{28}},
  \bibinfo{pages}{6904} (\bibinfo{year}{1983}).

\end{thebibliography}

\end{document}